# Increasing the Rate of Magnesium Intercalation Underneath Epitaxial Graphene on 6H-SiC(0001)

*Jimmy C. Kotsakidis[1*], Marc Currie[2], Antonija Grubišić-Čabo[1], Anton Tadich[3], Rachael L. Myers-Ward[2], Matthew DeJarld[2], Kevin M. Daniels[4], Chang Liu[1], Mark T. Edmonds[1], Amadeo L. Vázquez de Parga[5, 6], Michael S. Fuhrer[1], D. Kurt Gaskill[4*]*

[1] School of Physics and Astronomy, Monash University, *Melbourne, Victoria 3800, Australia.*

[2] U.S. Naval Research Laboratory, *Washington D.C. 20375, USA.*

[3] Australian Synchrotron, *800 Blackburn Rd, Melbourne, Victoria 3168, Australia.*

[4] Institute for Research in Electronics and Applied Physics, University of Maryland, *College Park, MD, 20742, USA*

[5] Dep. Física de la Materia Condensada and Condensed Matter Physics Center (IFIMAC), Universidad *Autónoma de Madrid, Cantoblanco 28049, Madrid, Spain.*

[6] IMDEA *Nanociencia, Cantoblanco 28049, Madrid, Spain.*

*Address correspondence to **jckot1@student.monash.edu** or **kurtcapt87@verizon.net**

## Abstract

Magnesium intercalated 'quasi-freestanding' bilayer graphene on 6H-SiC(0001) (Mg-QFSBLG) has many favorable properties (*e.g.*, highly n-type doped, relatively stable in ambient conditions). However, intercalation of Mg underneath monolayer graphene is challenging, requiring multiple intercalation steps. Here, we overcome these challenges and subsequently increase the rate of Mg intercalation by laser patterning (ablating) the graphene to form micron-sized discontinuities. We then use low energy electron diffraction to verify Mg-intercalation and conversion to Mg-QFSBLG, and X-ray photoelectron spectroscopy to determine the Mg intercalation rate for patterned and non-patterned samples. By modeling Mg intercalation with the Verhulst equation, we find that the intercalation rate increase for the patterned sample is 4.5±1.7. Since the edge length of the patterned sample is ≈5.2 times that of the non-patterned sample, the model implies that the increased intercalation rate is proportional to the increase in edge length. Moreover, Mg intercalation likely begins at graphene discontinuities in pristine samples (not step edges or flat terraces), where the 2D-like crystal growth of Mg-silicide proceeds. Our laser patterning technique may enable the rapid intercalation of other atomic or molecular species, thereby expanding upon the library of intercalants used to modify the characteristics of graphene, or other 2D materials and heterostructures.

## Introduction

Since the discovery and electrical measurement of graphene[1-2] – a two-dimensional (2D) plane of carbon atoms arranged in a hexagonal lattice structure – there has been extensive use of graphene in condensed matter physics experiments due to its ease of production (typically *via* micromechanical cleavage)[1-2] and access to novel physics owing to its linear 'Dirac' band structure at low energy.[3-4] But for graphene to be useful in device applications, we must be able to *engineer*[5] its electrical properties. To achieve this, various techniques such as surface chemical doping[6] or atom substitution (see refs. [5, 7] for a brief review of these methods), gating,[8-13] proximity effects,[8, 14] stacking[15-16] and intercalation[17-25] (see ref. [26] for a brief review of intercalation) have been applied. Of these techniques, intercalation (*i.e.,* the insertion of atoms or molecules beneath or in-between graphene sheets) has been demonstrated as a relatively simple and versatile method for achieving control over graphene's key electrical properties – carrier type,[17-19, 21] conductivity[20, 27-29] and bandgap[25, 30]. For instance, engineering the carrier type and bandgap *via* intercalation could enable graphene p-n junctions[31-32] and transistors.[33] Whereas increasing graphene's conductivity *via* intercalation could enable highly conductive and transparent electrodes for solar cells,[34] interconnects[35] or new superconductors.[20, 36]

Many of the intercalation experiments to date have utilized epitaxially synthesized graphene on SiC (either the 4H- or 6H- polytype on the (0001) orientation, *i.e.*, silicon face).[17-18, 20, 36-38] Graphene synthesis on SiC occurs because the higher vapor pressure of Si results in its preferential sublimation over C, and at high temperatures, the remaining C on SiC surface reorganizes into graphene.[39-40] One reason for the widespread use of epitaxial graphene on SiC in intercalation experiments is that the graphene is single crystalline and continuous over the SiC wafer surface. The resulting large area graphene permits surface sensitive characterization techniques such as X-ray photoelectron spectroscopy (XPS), or low energy electron diffraction (LEED) because the sampling areas for these techniques are typically much larger (≈10,000 μm$^2$) than those of graphene flakes produced by micro mechanical cleavage (≈100 μm$^2$). Thus far, ≈ 30 elements have intercalated graphene on SiC, and in almost every case, intercalation occurs solely below the 'buffer layer'[41] region (also referred to the 'interfacial' or 'zero layer' region). The buffer layer is a carbon layer that forms between the graphene and the SiC surface during growth. It is physically similar to graphene, but highly electrically resistive due to partial covalent bonding of the C to the topmost Si at the SiC surface.[42] During intercalation underneath the buffer layer, the intercalant breaks the covalent bonds between the buffer layer and SiC surface. The intercalant then typically bonds directly with the Si surface, liberating the buffer layer and forming a layer of 'quasi-freestanding'[17] graphene.[41] The intercalation process is usually conducted in an ultra-high vacuum (UHV) environment, but for higher vapor pressure elements such as hydrogen and antimony, ambient pressures (≈1 atm) have been implemented.[17, 43] The intercalation mechanism for graphene (whether on SiC, or any other substrate) generally falls into one of a few categories:[44] intercalation through large (multi-atomic) graphene voids, *i.e.*, discontinuities in the graphene,[31, 45-47] wrinkles with nano-sized defects,[48-49] existing point defects[50-52] or defects created by the intercalant itself.[32, 47, 53-54] In the case of graphene on SiC, intercalation has been observed to begin at the step edges[32, 55-56] (where ripples are thought to increase reactivity of the graphene)[57] and through terraces, as in the case of hydrogen.[21]

Recently, magnesium, despite having a higher vapor pressure than antimony, was able to be intercalated in UHV underneath epitaxial monolayer graphene (EMLG) on 6H-SiC(0001). This created a highly n-type doped (≈ 2 × 10$^{14}$ cm$^{-2}$)[24-25] and 'quasi-freestanding' bilayer graphene (Mg-QFSBLG) with a sizeable bandgap of ≈0.36 eV at the Dirac point.[25] As expected, the Mg intercalated between the buffer layer and the SiC surface to form a Mg-silicide type compound. (The formation of a metal silicide at the surface of the SiC after intercalation has been shown to occur with many other elements).[55, 58-65] Importantly, it was found that Mg-QFSBLG was relatively stable in ambient conditions for six hours.[24] Additional measurements showed that Mg was unable to intercalate hydrogen-intercalated quasi-freestanding bilayer graphene on 6H-SiC(0001)[24] – opening up the attractive possibility of a hybrid Mg/H intercalation procedure to create localized n-p type junctions. To achieve intercalation, Mg was evaporated onto the graphene, and then the sample was annealed. However, for successful Mg intercalation it was found that multiple intercalation steps were necessary.[24] The difficulty of Mg intercalation is potentially due to its high vapor pressure.[66] Furthermore, multiple deposition steps can increase surface contamination on the samples (*via* oxidation of the residual surface Mg), which is then difficult to remove by annealing after the sample is intercalated. This is due, in part, to the high sublimation temperature of magnesium oxide,[67] making its removal difficult or impossible without risk of de-intercalating the sample or destroying the graphene.[68] Thus, although Mg is a desirable intercalant – since it results in a highly n-type doped, large band gap bilayer graphene with low reactivity in ambient atmosphere and co-intercalant capability; its difficulty to intercalate underneath epitaxial graphene on SiC remains an engineering obstacle.

In this paper, we show how one can overcome this engineering obstacle and significantly increase the speed and efficiency at which Mg intercalates epitaxial graphene on SiC. In order to demonstrate this, we simultaneously intercalate (in a stepwise process consisting of 3 separate intercalations) two identically grown epitaxial monolayer graphene (EMLG) samples with Mg – one as-grown (*i.e.*, a standard EMLG sample), and another that we engineer with patterned lines defining areas with no graphene, that we call the EMLG-p sample. The patterned lines, which number 13 in total, have dimensions of ≈ 2 × 2600 μm$^2$ and are separated by ≈ 200 μm. A femtosecond pulsed laser is successfully used to locally pattern these lines in the graphene of a standard EMLG sample, as described previously (see also Experimental Overview and Methods).[69] We first analyze the samples using LEED, where we rapidly determine Mg-intercalation of the EMLG-p and EMLG samples taking place. Our LEED results show qualitatively that the EMLG-p sample intercalates at a faster rate than the EMLG sample. Next, we implement XPS to quantitatively measure the intercalation rate using the intensity of the SiC-related peaks in the C 1s spectra for each sample. Both LEED and XPS corroborate that the EMLG-p sample shows evidence of significant Mg-intercalation (*i.e.*, conversion to Mg-QFSBLG) after only the 1st intercalation step, whereas the standard EMLG sample only shows significant conversion to Mg-QFSBLG after the 3rd intercalation step. We fit a line to our data using a logistic function based on a solution to the Verhulst equation, subject to carrying capacity boundary conditions (conditions consistent with crystal growth in a finite area sample, *i.e.*, the physical scenario of our intercalation). We find that our data is well fit by this model, which shows an intercalation rate increase by a factor of 4.5 ± 1.7 for the EMLG-p sample, with respect to the standard EMLG sample. This value agrees with the value of ≈5.2, which is the predicted rate increase of the EMLG-p sample when considering that intercalation rate is proportional to the increased edge length due to patterning. We note here that throughout the text, we use the word 'rate' to describe the amount of Mg intercalation per intercalation step.

Our results strongly imply that the primary intercalation pathway for Mg is via discontinuities (*i.e.*, large multi-atomic holes) already present in the graphene prior to intercalation. Furthermore, the difference in intercalation onset between the two samples is surprising given the relatively low density of patterned lines. This suggests that pristine samples are being intercalated via discontinuities such as the edges of the sample, or scratches on the EMLG surface. In addition, we find that our data supports the Mg intercalant forming a '2D' silicide with the Si surface of the SiC – in agreement with our previous results.[24] Because our intercalation is gradual in the EMLG sample, we find that this silicide is continuously crystalline, suggesting that the intercalation mechanism proceeds via a crystal growth process. Nonetheless, our work demonstrates that the patterning of well-defined graphene discontinuities in EMLG samples can increase the rate and efficiency of Mg intercalation, and moreover, could enable the intercalation of other difficult-to-intercalate atoms or molecules – the properties of which are yet to be discovered.

## Experimental Overview

All graphene samples used in the experiment are monolayer graphene samples epitaxially synthesized on 6H-SiC(0001), the details of which are published elsewhere.[70] Briefly, the 6H-SiC(0001) substrates (II-VI Inc.) are semi-insulating, with a ≈0.1° miscut. The graphene is grown in a Aixtron/Epigress VP508 horizontal hot wall reactor, resulting in majority monolayer growth on the terraces of the SiC (see Methods for further details).[71] After graphene synthesis, half of the (sister) samples were patterned using a femtosecond pulsed laser (amplified Ti:sapphire laser at 800 nm wavelength producing 50 fs pulses at a 250 kHz rate), and 50× objective lens to form a ≈2 μm diameter focal spot. A total of 13 lines separated by ≈200 μm were patterned in the sample (see Figure 1a), in which the sample stage was moved from left to right at a rate between 20 – 40 μm s$^{-1}$. This graphene laser patterning technique has been successfully implemented, and is discussed in detail elsewhere[69] (see Methods for pertinent details), and results in the ablation (complete removal) of the graphene in these patterned areas (see Supporting Information, Figure S0). We call these samples patterned EMLG samples, or EMLG-p for short. All Mg intercalation experiments were carried out at the Australian Synchrotron soft X-ray beamline end station. The samples (EMLG, EMLG-p and H-QFSBLG-p samples – see below) were mounted on the same sample holder and Mg intercalation was performed in a stepwise fashion. The samples were characterized with LEED and XPS prior to, and after each Mg intercalation.

In addition to the EMLG-p sample, the same patterning procedure was also applied to a hydrogen intercalated quasi-freestanding bilayer graphene sample (H-QFBSLG-p), which was intercalated at the same time as the EMLG and EMLG-p samples. We did not find any evidence of Mg-intercalation for the H-QFSBLG-p sample, a conclusion identical to our previous work on standard H-QFSBLG samples.[24] Nonetheless, the patterned H-QFSBLG sample was useful in determining the identity of several peaks in the Si 2p, Mg 2p and O 1s core levels for the EMLG-p sample, the details of which are presented in the Supporting Information.

## Results and Discussion

We first overview our LEED results of both the EMLG and EMLG-p samples prior to Mg intercalation to establish the sample quality. Figure 1 shows a schematic overview of the EMLG-p sample, as well as a comparison of the LEED between the EMLG-p and EMLG sample. The schematic of the EMLG-p sample in Figure 1a shows a total of 13 patterned lines with dimensions ≈2 × 2600 μm$^2$ (separated by ≈200 μm), centered in the middle of the EMLG-p sample. These lines were patterned by application of a femtosecond pulsed laser. In these patterned regions, the graphene was completely ablated, leaving behind a bare 6H-SiC(0001) surface devoid of graphene (see Supporting Information Figure S1 for Raman spectroscopic analysis of the EMLG-p sample); effectively creating a higher density of 'edge' regions in the graphene. The inset in Figure 1a shows a magnified optical micrograph of these patterned regions. Figure 1b-c shows the subsequent LEED spot pattern of the EMLG-p and EMLG sample at an electron energy of 100 eV, respectively. Upon comparison, we observe no significant changes between the EMLG and EMLG-p samples, demonstrating that our patterning procedure does not significantly affect the graphene outside the patterned regions. For instance, both Figure 1b-c show G(1×1) diffraction spots corresponding to graphene, SiC(1×1) diffraction spots corresponding to the symmetry of the underlying SiC, and (6√3×6√3)R30° / (6×6) diffraction spots which arise from the buffer layer.[72]

The intercalation mechanism for graphene on various substrates has been found to occur via different processes, which we categorize here as either: intercalation occurring from already present defect features (*i.e.*, point defects, multi-atomic vacancies, or edges of the graphene), or from defect features created by the intercalant (*i.e.*, formation of mono/bi-vacancies or wrinkles with nano-sized atomic vacancies).[44] Since the vapor pressure of Mg is high,[66] and its affinity for carbon low,[73] we hypothesize that Mg intercalates through existing defect features already present. Thus, if intercalation begins from already existing defect features such as large multi-atomic vacancies or graphene edges, engineering more of these discontinuities in the graphene should provide easier access for the Mg intercalant. Alternatively, if intercalation proceeds *via* defect creation by the intercalant itself (on SiC terraces or step edges, for example),[32, 55] the engineered graphene discontinuities should not influence the intercalation process. To evaluate the influence of our engineered defects regarding Mg intercalation, we employed LEED and XPS during the stepwise intercalation process, the results of which are shown in Figures 2 – 4.

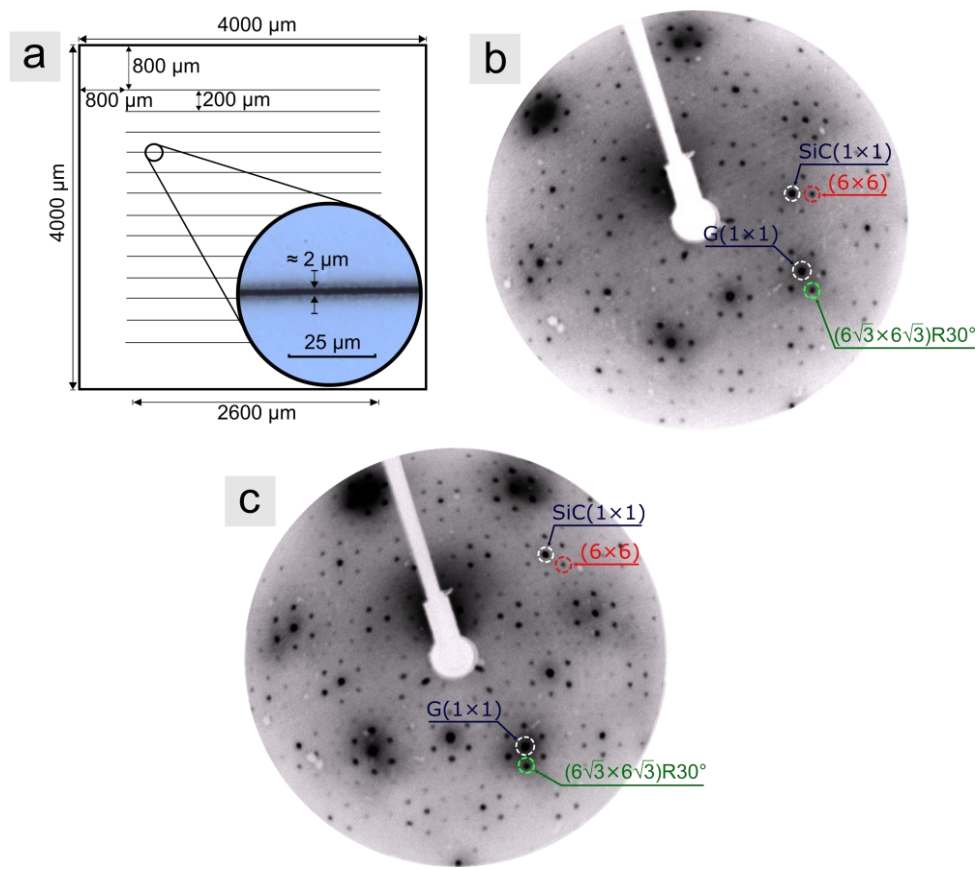

**Figure 1** – Overview of the samples prior to intercalation. **(a)** Schematic of the patterned epitaxial monolayer graphene (EMLG-p) sample showing (solid) lines with a width × length of ≈2 × 2600 µm representing where the graphene was ablated. Inset: magnified optical micrograph of the patterned line. Both **(b)** (EMLG-p) and **(c)** (EMLG) show low energy electron diffraction (LEED) spot patterns at an electron energy of 100 eV with graphene (G(1×1)), silicon carbide (SiC(1×1)) and buffer layer (6√3×6√3R30°) / (6×6) spots characteristic of a standard EMLG sample. Other than a trivial sample rotational difference, the LEED spot patterns of both EMLG-p and EMLG samples are identical.

We begin by examining the stepwise Mg intercalation of the EMLG-p sample with LEED. Figure 2 shows the LEED patterns of the EMLG-p after a stepwise Mg intercalation taken at an electron energy of 100 eV. Figure 2a shows the EMLG-p sample after the 1$^{st}$ Mg-intercalation. From the Figure 2a inset – which shows a magnified and contrast-enhanced view of the central part of the LEED pattern – it is clear that after the 1$^{st}$ Mg intercalation, several diffraction spots with a (√3×√3)R30° symmetry around the SiC(1×1) spots become apparent. These diffraction spots, which we label SiC(√3×√3)R30°, are related to the formation of an ordered (with respect to the SiC) Mg-silicide layer underneath the buffer layer of the EMLG-p sample to form Mg-QFSBLG.[24-25] Thus, our LEED results in Figure 2a imply significant conversion of EMLG-p to Mg-QFSBLG after only the 1$^{st}$ Mg intercalation. As for the coordination of the Mg atoms, recent modeling of the Mg-QFSBLG structure (based also upon previous LEED results),[24] has found that the Mg atoms on the surface of the SiC likely forms a unit cell with a ratio of Mg:Si of 1:3.[25]

After the 2$^{nd}$ Mg intercalation shown in Figure 2b, the intensity of the SiC(√3×√3)R30° diffraction spots increases, implying that more Mg has intercalated underneath the buffer layer to form Mg-silicide. The creation of a bilayer graphene sample (Mg-QFSBLG) is further supported by the increase in intensity of the G(1×1) diffraction spots with respect to the SiC(1×1) diffraction spots, and a decrease in the intensity of the buffer layer related (6√3×6√3)R30° / (6×6) diffraction spots. These observations are consistent with our recent study of Mg intercalated EMLG samples,[24-25] and also, with other SiC intercalation studies using H, Bi, In and Si.[41, 46, 74-75]

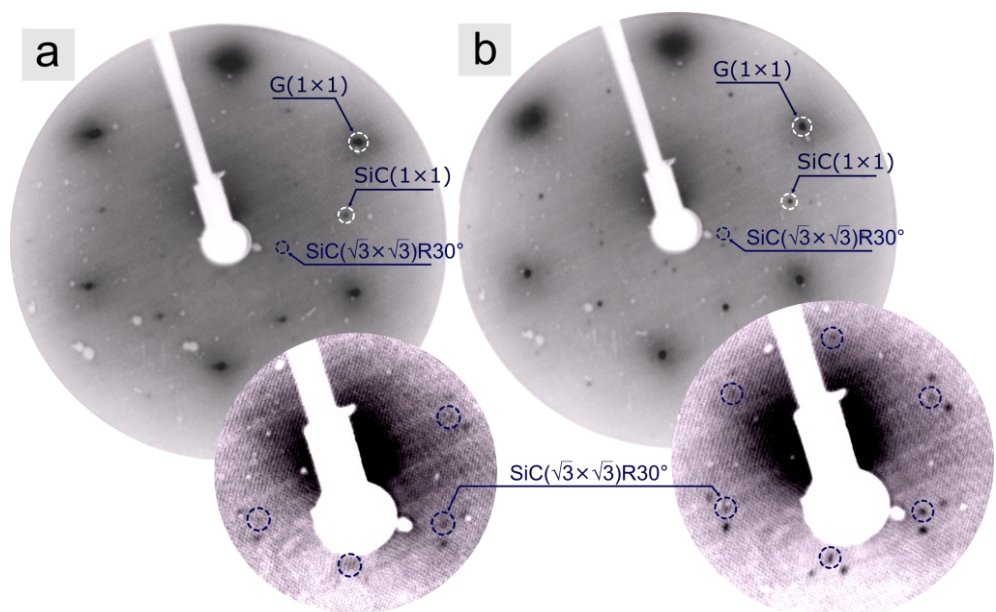

**Figure 2** – LEED patterns at an electron energy of 100 eV, showing changes in the EMLG-p sample after the **(a)** 1st and **(b)** 2nd Mg intercalation. Both (a) and (b), show the existence of SiC(√3×√3)R30° diffraction spots. These spots are strongly correlated to the formation of crystallographic Mg-silicide layer. This Mg-silicide layer develops underneath the buffer layer creating a Mg-intercalated 'quasi-freestanding' bilayer graphene *i.e.*, Mg-QFSBLG. Insets show a magnified and contrast-enhanced view of the central part of the LEED pattern shown in both (a) and (b).

Next, we examine the standard EMLG sample after the exact same stepwise Mg intercalation as the EMLG-p sample shown in Figure 2. Figure 3 shows the LEED patterns of the standard EMLG sample after each Mg intercalation step. As is evident in Figure 3a which shows the 1st Mg-intercalation step, no significant Mg-intercalation is observed from the LEED results. The LEED pattern is almost identical to the standard EMLG sample shown in Figure 1c. (The apparent 'blurring' present in Figure 3a, and Figure 2a, is due to excess Mg present on the surface after annealing – see Methods and Mg 2p core levels in Figures S4 and S7 in the Supporting Information). Hence, in Figure 3a we observe prominent (6√3×6√3)R30° / (6×6) diffraction spots, implying that there still exists a significant amount of buffer layer. Furthermore, there does not exist any observable SiC(√3×√3)R30° diffraction spots arising from an ordered Mg-silicide (to within the detection capabilities of our LEED). Thus, we conclude from the LEED data that there is no significant intercalation or conversion from EMLG to Mg-QFSBLG after the 1st Mg intercalation step.

But with further Mg intercalation, as in Figure 3b which shows the standard EMLG sample after the 2nd Mg intercalation step, there is evidence of Mg intercalation because of the presence of SiC(√3×√3)R30° diffraction spots. The inset of Figure 3b shows a magnified and contrast enhanced view of the central part of the LEED pattern. Under these conditions, we can observe faint SiC(√3×√3)R30° diffraction spots. Because we also observe distinct (6√3×6√3)R30° / (6×6) diffraction spots, this suggests that the sample is beginning to transition from EMLG to Mg-QFSBLG. Moreover, in both Figure 3a and 3b, there is only a small intensity difference between the G(1×1) and SiC(1×1) spots, thereby implying the lack of a graphene bilayer and therefore, minimal Mg-intercalation.

Figure 3c shows the 3$^{rd}$ (and final) Mg-intercalation step. The (6√3×6√3)R30° / (6×6) diffraction spots are now more faint, and the intensity difference between the G(1×1) and SiC(1×1) spots is much greater. These observations imply that the buffer layer has disappeared, and a graphene bilayer has formed, respectively.[24-25] Furthermore, the SiC(√3×√3)R30° diffraction spots which are related to the concurrent formation of the Mg-silicide, are much more distinct, in agreement with these observations. The results presented in Figure 3 are in stark contrast to the observations in Figure 2 for the EMLG-p sample, which showed evidence of intercalation and conversion to Mg-QFSBLG already at the 1$^{st}$ Mg-intercalation step.

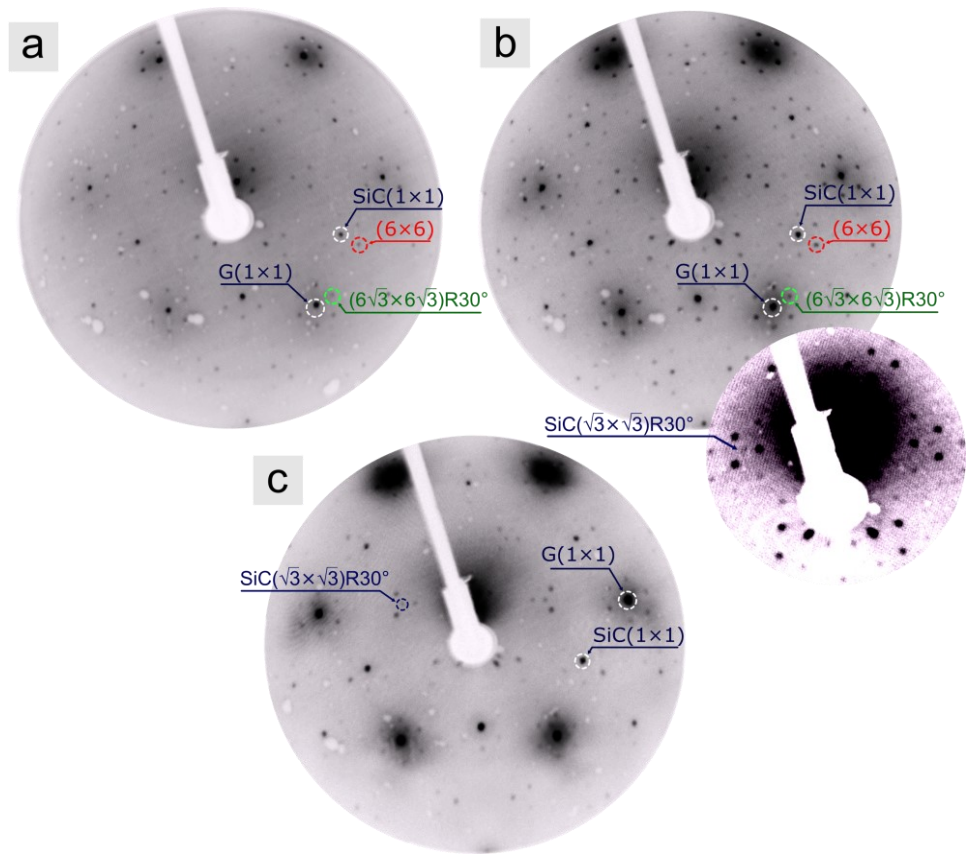

**Figure 3** – LEED patterns at an electron energy of 100 eV, showing the standard EMLG sample after a stepwise Mg intercalation. **(a)** 1$^{st}$ Mg-intercalation showing the presence of the buffer layer - (6√3×6√3)R30° / (6×6) spots, and similar intensity of the G(1×1) and SiC(1×1) spots implying no significant Mg intercalation. **(b)** 2$^{nd}$ Mg-intercalation showing the presence of the buffer layer - (6√3×6√3)R30° / (6×6) spots, and similar intensity of the G(1×1) and SiC(1×1) spots. Inset shows contrast enhanced and magnified view of central region and SiC(√3×√3)R30° diffraction spots, implying sample is in the process of transitioning to Mg-QFSBLG. **(c)** 3$^{rd}$ Mg-intercalation showing the presence of the SiC(√3×√3)R30° diffraction spots, large intensity discrepancy between the G(1×1) and SiC(1×1) spots and diminished intensity of the (6√3×6√3)R30° / (6×6) spots implying Mg-intercalation has occurred to form Mg-QFSBLG.

While LEED is useful in qualitatively determining whether Mg-intercalation has occurred, the stepwise chemical and structural rate change of the samples are more accurately determined using XPS. This is illustrated in Figure 4a-b, which show a waterfall plot of the C 1s core level for each Mg intercalation step (1st Mg-intercalation, 'int. 1' *etc.*) of both the standard EMLG sample (Figure 4a) and the EMLG-p sample (Figure 4b). In these plots, the baseline corrected XPS data (see Methods) is shown as hollow circles, and the total fit to the data is given by the red (solid) line. We note here that the Si 2p and Mg 2p spectra for the EMLG and EMLG-p samples are shown in the Supporting Information, and corroborate our findings here (*i.e.*, both show the formation of a Mg-silicide). Furthermore, all component binding energies ($E_B$) quoted in the main text have been averaged with all available values (from both the EMLG and EMLG-p spectra) – see Methods for details.

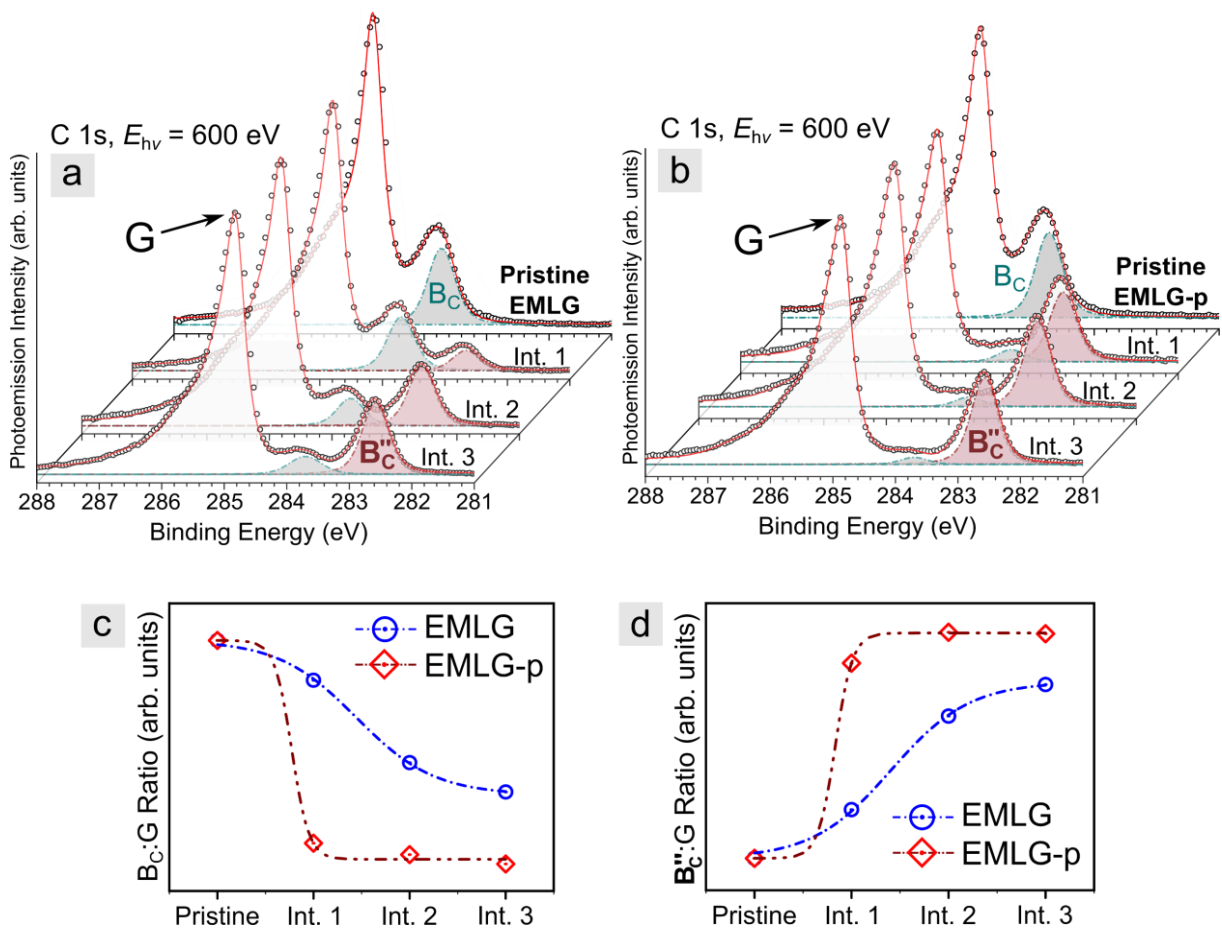

**Figure 4** – XPS measurements of the C 1s photoelectron spectra for both the standard epitaxial monolayer graphene (EMLG) and patterned EMLG (EMLG-p) samples at an incident X-ray energy of $E_{h\nu}$ = 600 eV. **(a)** Standard EMLG sample shows the evolution of the C 1s spectrum as a function of Mg-intercalation step. Component $B_C$ corresponds to the carbon in the SiC bulk and component $B_C''$ corresponds to the carbon in the SiC bulk *after* Mg-intercalation. **(b)** Showing the evolution of the C 1s spectrum as a function of Mg-intercalation step for the EMLG-p sample. **(c)** The $B_C$:G ratio versus intercalation step decreases as Mg-intercalation progresses. **(d)** The $B_C''$:G ratio versus intercalation step increases as intercalation progresses. The blue (circles) and red (diamonds) curves represent logistic function fits (Equation 3) to the Verhulst model (Equation 2) for the EMLG and EMLG-p data, respectively. The fits indicate an increase in the rate of intercalation for the EMLG-p sample by 4.5 ± 1.7 for both the $B_C$:G and $B_C''$:G data.

The most important features of the fitted line shown in Figure 4a-b are captured by analysis of components $B_C$ and $\bm{B_C''}$. Components $B_C$ (at 283.69 ± 0.05 eV) and $\bm{B_C''}$ (at 282.61 ± 0.10 eV) correspond to the carbon atoms in the first few layers of the SiC bulk prior to Mg intercalation (*i.e.*, the pristine sample) and after Mg intercalation, respectively (we find no evidence for the formation of a magnesium carbide – see Supporting Information, Section 6 for further discussion).[24] Relating the change in the intensity of the SiC peak (*i.e.*, $B_C$ to $\bm{B_C''}$) in the C 1s spectrum, to the magnitude of intercalation is not a new analysis technique, and various examples can be found in literature.[31, 42, 75-76] Thus, we expect that as the sample becomes more intercalated, component $B_C$ should diminish in photoemission intensity, and at the same time, component $\bm{B_C''}$ (located at lower binding energy) should increase in photoemission intensity relative to the graphene peak at 284.83 ± 0.05 eV (labelled as 'G' in Figure 4a-b, see the Supporting Information for further details on the graphene fitting procedure). That is, $B_C$ and $\bm{B_C''}$ should possess an inverse relationship of the form:

$$[B_C]_n + [\bm{B_C''}]_n = B_0 \tag{1}$$

Where $[B_C]_n$ and $[\bm{B_C''}]_n$ represent the concentration (intensity) of component $B_C$ and $\bm{B_C''}$ at intercalation step *n*, respectively, and $B_0$ represents the initial intensity of the $B_C$ component at *n* = 0 (pristine sample prior to intercalation).

Hence, by plotting the intensity ratio ($B_C$:G or $\bm{B_C''}$:G) as a function of intercalation step, we can determine the 'rate' at which each sample intercalates in a more quantitative way than LEED, and this is displayed in Figure 4c (for $B_C$:G) and Figure 4d (for $\bm{B_C''}$:G). To fit the data in Figure 4c-d, we used the first order non-linear ordinary differential equation of the form:[77-78]

$$\frac{dB}{dn} = rB\left(1 - \frac{B}{K}\right), \tag{2}$$

where *r* is the growth rate and *K* is the 'carrying capacity' or value which the expression approaches asymptotically. Equation 2 is a form of the Verhulst equation,[79-81] which describes the rate change of a quantity, subject to a 'carrying capacity' or maximum amount of permissible growth. Equation 2 can be solved exactly to a function of the form:

$$B(n) = \frac{K}{1 + ae^{-rn}} \tag{3}$$

Where *n* is the intercalation step (*i.e.*, *n* = 0, 1, 2, 3), *K* is a constant (≥ 0) that defines the final steady state value, *a* is a positive constant which translates the curve in the x-axis (see Supporting Information for further details), and *r* is a rate constant that defines the rate of change between the initial and final steady state value. In the literature, the function defined by Equation 3 is called the logistic function.[79] The logistic function has been used extensively in crystal nucleation and growth modeling to represent autocatalytic reactions.[82-86] Equation 3 was used to fit the data in Figures 4c-d using a damped least squares fitting procedure (see Methods).

Figure 4c shows that component $B_C$ (carbon atoms in the bulk of pristine SiC, see above) decreases in photoemission intensity as the sample becomes more Mg-intercalated, while component $\bm{B_C''}$ simultaneously increases. But the rate at which this occurs is different for the EMLG and EMLG-p sample. Using Equation 3, the rate at which component $B_C$ decreases per intercalation step (with respect to component G) is $r \approx 2.47$ for the EMLG sample, whereas component $B_C$ decreases at a rate of $r \approx 11.18$ per intercalation step for the EMLG-p sample. On the other hand, Figure 4d shows that component $B_C''$ increases in photoemission intensity as the sample becomes more Mg-intercalated. The rate at which this occurs is found to be $r \approx 2.48$ for the EMLG sample, and $r \approx 11.18$ for the EMLG-p sample. This result is expected, since the rate constant of component $B_C''$ should vary oppositely to component $B_C$, consistent with Equation 1. The goodness of fit in Figures 4c-d (Adj. R-square > 0.99 for all fitted lines – see Supporting Information) further strengthens this claim. The increase in rate, $r$, from EMLG to EMLG-p is thus, $\approx 11.18/2.48 = 4.5 \pm 1.7$. The large uncertainty in this value arises from the fact that there exists no data point between Pristine and Int. 1 for EMLG-p, and so the rate constant could vary between $r = 7 - 15$ (see Supporting Information for further details regarding uncertainties).

At the beginning of this section, we noted that if intercalation proceeds via intercalant defect creation (through step edges or terraces for example), then there should be no difference in the rate of Mg intercalation between EMLG and EMLG-p; experimentally this is clearly not the case (see Fig. 4c-d) and so we rule out this mechanism. Alternatively, if the intercalation mechanism is dependent upon defect features such as graphene edges, engineering more of these discontinuities in the graphene should enhance the rate of Mg intercalation, which we do observe experimentally. We used Equation 2 as it models well the physical processes of the intercalation mechanism we propose. That is, initially, the Mg will form crystal seeds of the (2D) Mg-silicide type compound governed by some initial rate. Once formed, these Mg-silicide seed crystals grow in the same orientation (as evidenced by our LEED results in Figs. 2-3) and in a rapid manner (*i.e.*, faster than the initial rate of seed crystallization). In the case of the EMLG-p sample, this should occur faster over the standard EMLG sample for two reasons. Firstly, the increased number of edge sites in the EMLG-p sample will necessarily enable the formation of more crystal seeds. Secondly, if edge intercalation from graphene discontinuities is favored, then intercalation in the standard EMLG sample is progressing from the outside in, and the amount of Mg intercalated is competing with sublimation of the Mg from the surface of the sample. By opening more pathways for Mg to intercalate in the center of the sample, more Mg can intercalate before it is sublimated. The intercalation ends when the surface of the SiC becomes completely populated by Mg, or the Mg is exhausted (growth slows and is limited due to carrying capacity).

Although the model fits the XPS data well, we can further test the model in another way. Since the intercalation of Mg begins at the edges of the graphene, and these permit access of the Mg to the surface of the SiC; then the change in intercalation rate (given by the ratio of rate constants found above) should be proportional to the difference in total edge length between the EMLG and EMLG-p sample. We find that the total edge length of the EMLG-p and EMLG samples are 16,000 µm and 83,704 µm, respectively, yielding a ratio of $\approx 5.2$. Our experimental ratio of $4.5 \pm 1.7$ agrees with this theoretical value, and thus, implies that our observed intercalation rate increase is proportional to the increased edge length of the EMLG-p sample.

While the logistic function in Equation 3 has been used by others to model the growth kinetics of crystals,[82-86] and while our fit to the data is good (see Supporting Information), logistic models are notoriously difficult to interpret and assign a physical basis.[87] Thus, we believe that further work is necessary to fully assess the model proposed here. For instance, further investigations with a low energy electron microscope (LEEM) should corroborate our findings. As a further aid, we have included another commonly used kinetic model (the Kolmogorov-Johnson-Mehl-Avrami KJMA, or JMAK model)[88-90] in the Supporting Information. However, we find that this model does not fit our data as well as the logistic function (Equation 3), nor does it yield a rate constant that is proportional to the increased edge length of the EMLG-p sample. What is clear is that Figure 4 demonstrates that the EMLG-p sample intercalates at a faster rate than the standard EMLG sample, yielding a solution to the Mg intercalation engineering problem. Moreover, the model implies that the EMLG-p sample reaches the saturation point (*i.e.*, fully intercalated) after the 3$^{rd}$ intercalation step, whereas the EMLG sample does not.

Our proposed Mg intercalation model also offers a unique insight into the mechanics of the intercalation process itself. In general, Mg intercalates underneath the buffer layer, forms a silicide with the Si at the SiC(0001) surface and consequently, breaks the C sp$^3$ bonds of the buffer layer. The buffer layer subsequently becomes a second layer of graphene.[24-25] Since intercalation rate increases proportionally with edge length (increased graphene discontinuities), this implies the suppressed intercalation efficiency of standard "pristine" EMLG samples is likely due to significant reduction of Mg flux to the substrate surface, as there are minimal discontinuities in pristine graphene which provide pathways to the SiC(0001) surface (and so intercalation begins at the edges of the sample, far from the center). A reason for multi-atomic graphene discontinuities being the preferred site of intercalation rather than step edges or flat terraces, is that Mg likely lacks the 'reactivity'[47, 55, 73] with carbon (under the conditions of intercalation used here). Hence, it is likely that the Mg intercalation of pristine EMLG samples proceeds only because of the presence of multi-atomic discontinuities that naturally occur such as from the edges of the sample, small defects in the starting SiC substrate or mishandling of the sample (*i.e.,* scratches on the surface from tweezers or sample clamps).

Patterning EMLG samples *via* the removal of graphene to facilitate either localized intercalation, or easier access for the intercalation of specific atomic species, has been demonstrated previously. For example, Sorger *et al.*[31] used an oxygen plasma etch to define ≈200 nm diameter holes separated by ≈300, 500, 700 and 1000 nm in an EMLG sample, in order to fabricate localized p-n junctions using hydrogen as the intercalant. Here, it was found that separation distances of 1000 nm did not result in significant localized hydrogen intercalation. Moreover, Feldberg *et al.*[91] intercalated Ga and In bilayers underneath EMLG using a nitrogen plasma for defect creation to help facilitate intercalation, and showed that standard samples that had not undergone a nitrogen plasma step did not intercalate Ga and In significantly. More recently, Briggs *et al.*[68] used an oxygen/helium plasma to generate carbon vacancy defects in EMLG samples to help facilitate intercalation of (separately) Ga, Sn and In layers underneath the buffer layer (in a process dubbed 'confinement heteroepitaxy'). Although the intercalation of H, Ga, Sn and In underneath the buffer layer of EMLG has been demonstrated by others using non-engineered EMLG samples,[17, 32, 63, 75, 92] the engineering of defects (*i.e.*, multi-atomic discontinuities in the graphene) can significantly promote intercalation.[31, 68, 91]

The engineered defects in these previous works were of relatively high density, with a separation of ≈300-700 nm,[31] or directly observable in the C 1s spectra as judged with a lab-based XPS.[68] Our work differs from these previous efforts in that the patterning of the substrate does not necessarily need to be in such high density in order to greatly facilitate intercalation. Our patterned lines have a dimension of 2 × 2600 μm$^2$, and are separated by 200 μm, in stark contrast with previous efforts.[31, 68, 91] Our 'defect-free' areas are at least three orders of magnitude greater, and thus, have the benefit of large defect-free areas in which devices could be fabricated. Furthermore, we have demonstrated that the rate of Mg-intercalation increases proportionally with the increase in graphene edge length. This should further aid the modeling of Mg-intercalation (and elements that intercalate similarly) for device applications.

Moreover, we have demonstrated that the patterning method implemented here results in the significant intercalation of a high vapor pressure element after only a single intercalation step under UHV conditions. This is in contrast to the more sophisticated high-pressure Ar intercalation method for Sb employed recently by Wolff et al.[43]. Thus, it may be beneficial to attempt this method using other high vapor pressure elements such as Sb or Zn.[93] In addition, the single intercalation step should be beneficial in keeping the surface of the graphene free from contamination – which may arise as a direct consequence of depositing excess metal on the surface from multiple intercalation steps.

## Conclusion

Our LEED results verified Mg-intercalation in both EMLG and EMLG-p samples, and conversion of these samples into an Mg-intercalated quasi-freestanding bilayer graphene structure – Mg-QFSBLG – in accordance with previous results. The intercalation of Mg was able to proceed by penetrating underneath the buffer layer, forming a Mg-silicide type compound with the Si at the SiC(0001) surface. This process breaks the C $sp^3$ bonds of the buffer layer, which subsequently becomes a second layer of graphene. Continued Mg intercalation continues to form Mg-silicide that maintains a fixed orientation with respect to the SiC surface. Our LEED results already hinted that the EMLG-p sample intercalated at a faster rate than the standard EMLG sample, but we implemented XPS to quantitatively measure this rate. By application of the Verhulst equation, we were able to model the basic physical details of intercalation, and fit a logistic function to our data. The model predicts that the intercalation rate is proportional to the edge length, since intercalation should be seeded at multi-atomic graphene discontinuities (*i.e.*, edges of the graphene) and grow outwards from these points; also allowing a pathway for the Mg to flow to the SiC surface. This process of 2D-like crystal growth of the Mg-silicide continues until all the surface Si has reacted, or the Mg source has exhausted. The results imply that the preferred route of Mg intercalation is through multi-atomic graphene discontinuities, rather through step edges or flat terraces. Future intercalation work could also involve the use of LEEM and the Verhulst equation, but should simply replicate our main findings reported here.

Nonetheless, our results clearly demonstrate that the patterning technique employed can overcome the difficulties of intercalating Mg underneath EMLG. The significant rate increase of the EMLG-p sample over the EMLG sample is surprising, since our patterned lines are spaced at least three orders of magnitude further apart than in previous publications that used similar methods to facilitate intercalation. Moreover, this technique should be technologically relevant, since our patterned lines are spaced far enough to yield defect free spaces on the order of 100,000 $\mu m^2$, suitable for device fabrication.

Lastly, this model and the patterning approach used here may be applicable to other intercalated elements or molecules, subject to verification, as an aid to increasing intercalation rate. For example, both Sb and Zn are known to form silicides, which may be a requirement for our model to be applicable. Such an approach could further expand the library of intercalants which can engineer the properties of graphene.

# Methods

**Sample Growth:** All epitaxial monolayer graphene samples were synthesized on semi-insulating 6H-SiC(0001), nominally on-axis (≈ 0.1° miscut) substrates (II-VI Incorporated). The graphene synthesis procedure utilized in these experiments has been described previously.[24, 70] Briefly, the bare 6H-SiC(0001) substrates were initially etched in a Aixtron/Epigress VP508 horizontal hot wall reactor by flowing (laminar) approximately 5 standard liters per minute (slm) of Pd purified $H_2$ gas at 200 mbar, while the temperature was ramped (in ≈ 45 minutes) from 973 K to ≈ 1853 K. The H2 was switched to high purity Ar (10 slm), and pressure decreased to 100 mbar. Then, graphene synthesis took place for ≈ 20 minutes, resulting in the formation of nominally monolayer graphene (with small amounts of bilayer and trilayer graphene – see Supporting Information Figure S1). For H-QFSBLG samples, upon cooling to 1173 K, the 10 slm of Ar was switched to 80 slm of $H_2$ and hydrogen intercalation proceeded for 60 minutes at 900 mbar.

**Laser Patterning:** After graphene synthesis, an EMLG and H-QFSBLG sample were laser patterned using 50 fs (FWHM) optical pulses from an amplified Ti:sapphire laser at 800 nm wavelength and a 250 kHz rate. This same technique has been applied in a previous publication.[69] The laser light was directed through a 50× objective lens (≈5 mW measured after the objective lens), forming an approximate 2 μm focal spot on the surface of the sample. A total of 13 lines of length ≈2600 μm and width ≈2 μm, separated by ≈200 μm were patterned in both the EMLG and H-QFSBLG sample, in which the sample stage was scanned at a rate between 20 – 40 μm $s^{-1}$. There was a slight decrease in intensity of the laser from the extremities of the line towards the center.

**Sample Intercalation:** The samples were mounted on the same sample holder, side-by-side, with the temperature measured via a thermocouple in direct contact with the samples, and a single-color pyrometer. Prior to measurement and Mg-intercalation, the samples were cleaned by annealing in UHV ($1\times10^{-10}$ mbar) at ≈723 K for ≈5 hours. The surface oxide contamination after this annealing period was almost at the noise level of the measurement, and can be found in Supporting Information Figure S4. Magnesium (99.95 %, 1/8-inch turnings) was loaded into a NTEZ effusion cell (MBE Komponenten) equipped with a pyrolytic boron nitride (PBN) crucible. The samples were Mg-intercalated in a three-step process. In the $1^{st}$ Mg-intercalation step, ≈ 15 nm of Mg was deposited on the surface of the samples using a cell temperature of 673 K and deposition time of 20 minutes. The Mg film thickness was determined by a quartz crystal microbalance (Sycon Instruments$^{TM}$). The samples were then annealed at 463 K for 1.5 hours to drive off any excess Mg, and facilitate the intercalation process. Because this anneal temperature was lower in comparison with our previous work,[24] there was some metallic Mg residue left on the surface. This can be seen in our LEED images (Figure 2a and Figure 3a) which exhibit a blurring effect; and the Mg 2p spectra (see Supporting Information, Figures S3 and S7), which show significant metallic Mg present on the surface after the $1^{st}$ Mg-intercalation. Thus, in subsequent anneals, the temperature was increased to avoid excess Mg blurring of the LEED images and surface contamination, resulting in the removal of much of the metallic Mg. In the $2^{nd}$ Mg-intercalation step, a further ≈ 15 nm of Mg was deposited on the surface of the samples, which were then annealed at 623 K for 1.5 hours. In the $3^{rd}$ Mg-intercalation step, the $2^{nd}$ Mg-intercalation step was repeated. These temperatures are consistent with our previous work.[24] When the sample is annealed, some of the Mg will sublimate, and some will intercalate.

Thus, so long as the samples are intercalating, a comparison can be made about the efficacy of this process since both samples were mounted side-by-side and subjected to the same conditions (*i.e.*, the rate constant in Equation 3 could change, but the proportionality between EMLG and EMLG-p should be constant).

**Low Energy Electron Diffraction (LEED) and X-ray Photoelectron Spectroscopy (XPS):** All sample LEED and XPS measurements were taken at the Australian Synchrotron soft X-ray beamline. For LEED measurements, an 8-inch LEED spectrometer (OCI Vacuum Microengineering Inc.) was used. This instrument has a circular beam profile, approximately 200 µm in diameter. All measurements were taken at an electron energy of 100 eV. Initial measurements were taken on pristine (un-intercalated) samples prior to intercalation. The physical coordinates of these initial measurements were noted, and for each subsequent intercalation, the LEED measurements were taken at the same location (to within a 50 µm accuracy).

For XPS measurements, a SPECS PHOIBOS™ 150 electron spectrometer was used. The synchrotron beam spot size is approximately 80 × 120 µm$^2$, and roughly corresponds to the area measured by the detector. The pass energy of the analyzer was set at 10 eV for the O 1s core levels (with an energy step of 0.1 eV – see Supporting Information Figure S4), and 5 eV for all other core levels. Thus, the energy resolution in the spectra was estimated as ±0.05 eV for the C 1s, Si 2p and Mg 2p core level data. Each spectrum was calibrated in binding energy using the Au 4f$_{7/2}$ core level (reference binding energy = 84 eV) from an Au foil in electrical contact with the samples. Inelastic contributions from the data were removed by implementing a Shirley background subtraction. The deconvoluted spectra shown in the C 1s core levels in Figure 4 were fit with Voigt lineshapes, as were almost all other spectra (see Supporting Information), with the exception of the graphene related C 1s core level peak, which was fit with a Breit-Wigner-Fano (BWF) lineshape – in accordance with recently published work.[24] The details of the fitting procedure and overview of the deconvoluted lineshapes can be found in the Supporting Information, and is almost identical to recently published work.[24] The binding energy values for each component in the C 1s spectra (G, $B_C$, $B_C^{''}$) were calculated using the average value from Supporting Information Tables S2, S3, S6 and S7. In the case of the 'G' peak, the intercalated and non-intercalated positions were averaged (since there was no measured movement in the peak – see Supporting Information, Section 4 and 5). The uncertainty was then taken as the standard deviation, or ±0.05 eV, whichever was larger.

**Fitting of Figure 4c and 4c:** Components $B_C$ and $B_C^{''}$ were compared with the graphene peak 'G' to prevent uncertainties arising from excess Mg on the surface (see Mg 2p spectra in Supporting Information), causing an artificial attenuation of intensity. Furthermore, only the change in intensity from the pristine sample was taken, so that any intensity differences between the samples could be negated (*i.e.*, the change in $B_C$:G from pristine to Int.1 constituted 'Int. 1', not its absolute value). Once plotted, the data was fit using the logistic function shown in Equation 3 of the main text. We implemented a damped (Levenberg-Marquardt) least squares fitting procedure to fit all curves. In the case of the EMLG-p sample, the fitting procedure converged for the curve shown in Figure 4d ($B_C^{''}$:G). We then applied the same magnitude rate constant value, $|r|$, to the EMLG-p curve in Figure 4c describing the decrease in intensity of component $B_C$ to maintain fit consistency. The rationale behind this, and further information on the fit can be found in the Supporting Information.

# Supporting Information

The Supporting Information is available from the Wiley Online Library or from the corresponding authors.

The Supporting Information shows in Section 1 the Raman spectroscopic map of the EMLG-p sample after Mg intercalation (and exposure to ambient atmosphere for ≈8 months), demonstrating the spectral difference between the patterned and non-patterned regions of the sample. The mathematical models used to fit the XPS data and Figure 4c-d are shown in Section 2. Here we also briefly explain the Verhulst equation, the solution of which is a logistic function and compare this function to solution for another commonly used kinetic growth model – the Kolmogorov-Johnson-Mehl-Avrami (KJMA, or JMAK) model. Section 3 shows the table of values for all fitted components relating to the Si 2p and C 1s core level spectra for both EMLG and EMLG-p. The Si 2p, C 1s and Mg 2p spectra are then shown in Section 4, with all fitted components for the EMLG sample. Moreover, the Si 2p and C 1s spectra for the EMLG-p sample are shown in Section 5 with all fitted components. Finally, Section 6 shows the C 1s, Si 2p, Mg 2p and O 1s spectra for the H-QFSBLG-p sample, and the Mg 2p and O 1s spectra of H-QFSBLG-p are compared with the Mg 2p and O 1s spectra of EMLG-p.

**AUTHOR CONTRIBUTIONS**

J.C.K., M.C. and D.K.G. conceived the initial idea for the intercalation experiment using patterned samples. Samples were grown by D.K.G., R.L.M.-W., K.M.D. and M.D., and patterned by M.C. J.C.K. wrote the experimental proposal for the experiments at the Australian Synchrotron, with assistance from A.T., M.S.F and grammatical corrections from A.G.-C. Mg intercalation experiments were initially conceived by J.C.K, and previous intercalation recipes using Ca with A.L.V.P were extremely useful in determining intercalation conditions for Mg. Experimental results at the Australian Synchrotron were collected by J.C.K., A.G.-C., A.L.V.P., A.T., C.L., and M.T.E. Raman spectroscopy measurements were made by J.C.K with the assistance of D.K.G and R.L.M.-W. J.C.K analyzed all results, with assistance from D.K.G. and M.S.F. The manuscript was composed by J.C.K., with intellectual contributions from D.K.G. and M.C. The final manuscript was comprised through contributions given by all authors. All authors have given approval to the final version of the manuscript.

**FUNDING**

J.C.K. acknowledges the Australian Government Research Training Program and the Monash Centre for Atomically Thin Materials (MCATM) for financial support. J.C.K. and M.S.F. acknowledge funding support for the Australian Research Council (ARC) Laureate Fellowship (FL120100038) and the ARC Centre of Excellence in Future Low-Energy Electronics (CE170100039). A.L.V.P. acknowledges funding support from the Ministerio de Ciencia Innovatión y Universidades project PGC2018-093291-B-I00 and Comunidad de Madrid Project NMAT2D-CM P2018/NMT-4511. D.K.G., R.L.M.-W., M.D., K.M.D., and M.C. acknowledge support by core programs at the U.S. Naval Research Laboratory funded by the Office of Naval Research. This research was undertaken on the soft X-ray beamline at the Australian Synchrotron, part of ANSTO.

**CONFLICTS OF INTEREST**

The authors declare no conflicts of interest.

# Acknowledgements

J.C.K. acknowledges the hospitality of the U.S. Naval Research Laboratory (NRL) and for insightful discussions and use of equipment.

# Data Availability Statement

The data that supports the findings of this study are available in the supplementary material of this article.

# Abbreviations

EMLG, Epitaxial monolayer graphene on 6H-SiC(0001); EMLG-p, Patterned epitaxial monolayer graphene on 6H-SiC(0001); $E_B$, binding energy; $E_{h\nu}$, incident X-ray beam energy; H-QFSBLG, hydrogen intercalated 'quasi-freestanding' bilayer graphene on 6H-SiC(0001); LEED, low energy electron diffraction; LEEM, low energy electron microscopy; Mg-QFSBLG, magnesium intercalated 'quasi-freestanding' bilayer graphene on 6H-SiC(0001); XPS, X-ray photoelectron spectroscopy.

# Keywords

Magnesium, Intercalation, Graphene, Epitaxial Graphene, Silicon Carbide